\documentclass[10pt,letterpaper]{article}
\usepackage[top=0.85in,left=2.75in,footskip=0.75in]{geometry}

\usepackage{amsmath,amssymb}

\usepackage{changepage}

\usepackage[utf8x]{inputenc}

\usepackage{textcomp,marvosym}

\usepackage{cite}

\usepackage{nameref,hyperref}

\usepackage[right]{lineno}

\usepackage{microtype}
\DisableLigatures[f]{encoding = *, family = * }

\usepackage[table]{xcolor}

\usepackage{array}

\newcolumntype{+}{!{\vrule width 2pt}}

\newlength\savedwidth

\raggedright
\usepackage[aboveskip=1pt,labelfont=bf,labelsep=period,justification=raggedright,singlelinecheck=off]{caption}

\makeatletter
\renewcommand{\@biblabel}[1]{\quad#1.}
\makeatother

\usepackage{lastpage,fancyhdr,graphicx}
\usepackage{epstopdf}
\fancyheadoffset[L]{2.25in}
\fancyfootoffset[L]{2.25in}
\usepackage{setspace} 
\usepackage{float}
\newtheorem{theorem}{Theorem}

\newtheorem{definition}{Definition}
\newtheorem{example}[theorem]{Example}

\newtheorem{lemma}{Lemma}

\newtheorem{proposition}{Proposition}
\newtheorem{remark}{Remark}

\begin{document}
\vspace*{0.2in}

% Title must be 250 characters or less.
\begin{flushleft}
{\Large
\textbf\newline{A Model of Conflict and Leadership: Is There a Hawkish Drift in Politics?} % Please use "sentence case" for title and headings (capitalize only the first word in a title (or heading), the first word in a subtitle (or subheading), and any proper nouns).
}
\newline
% Insert author names, affiliations and corresponding author email (do not include titles, positions, or degrees).
\\
Siddhartha Bandyopadhyay \textsuperscript{1\Yinyang *},
Amit K Chattopadhyay\textsuperscript{2\Yinyang},
Mandar Oak\textsuperscript{3\Yinyang}
\\
\bigskip
\textbf{1} Affiliation Department of Economics, University of Birmingham, Birmingham B15 2TT, UK
\\
\textbf{2} Affiliation Department of Mathematics, Aston University, Birmingham B4 7ET, UK
\\
\textbf{3} Affiliation School of Economics, University of Adelaide, Adelaide SA 5005, AUSTRALIA
\\
\bigskip

% Insert additional author notes using the symbols described below. Insert symbol callouts after author names as necessary.
% 
% Remove or comment out the author notes below if they aren't used.
%
% Primary Equal Contribution Note
\Yinyang These authors contributed equally to this work.

% Additional Equal Contribution Note
% Also use this double-dagger symbol for special authorship notes, such as senior authorship.
%\ddag These authors also contributed equally to this work.

% Current address notes
%\textcurrency Current Address: Dept/Program/Center, Institution Name, City, State, Country % change symbol to "\textcurrency a" if more than one current address note
% \textcurrency b Insert second current address 
% \textcurrency c Insert third current address

% Deceased author note
%\dag Deceased

% Group/Consortium Author Note
%\textpilcrow Membership list can be found in the Acknowledgments section.

% Use the asterisk to denote corresponding authorship and provide email address in note below.
* s.bandyopadhyay@bham.ac.uk

\end{flushleft}
% Please keep the abstract below 300 words
\section*{Abstract}
We analyze conflict between a citizenry and an insurgent group over a fixed resource such as land. The citizenry has an elected leader who proposes a division such that, the lower the land ceded to the insurgents, the higher the cost of conflict. Leaders
differ in ability and ideology such that the higher the leader's ability, the lower the cost
of conflict, and the more hawkish the leader, the
higher his utility from retaining land. We show that the conflict arises from the political process with re-election motives causing leaders to choose to cede too little land to signal their ability.  We also show that when the rents of office are high, the political equilibrium and the second best diverge; in particular, the policy under the political equilibrium is more hawkish compared to the second best. When both ideology and ability are unknown, we provide a plausible condition under which the probability of re-election increases in the leader's hawkishness, thereby providing an explanation for why hawkish politicians may have a natural advantage under the electoral process.

%\linenumbers

% Use "Eq" instead of "Equation" for equation citations.
\section*{Introduction}\label{introduction section}
In this paper, we seek to understand the dynamics of conflict negotiation between a citizenry and an insurgent group in which the citizenry is represented by a politician who is subject to the political process. Specifically, we address the
following questions: When political leaders negotiating the
settlement face re-election prospects, do they tend to choose more extremist
(or hawkish, in a sense to be made more precise) policies? Do intrinsically more hawkish
politicians have an inherent re-election advantage as conflict becomes more salient? Does there exist a political failure in the sense that the equilibrium outcome of the political process fails to achieve a constrained Pareto optimal outcome (i.e., the best outcome given the information constraints, a.k.a. the second best)? And if so, how does the nature of this failure vary with the politician's preferences and ability?

There is at least casual evidence that during times of conflict hawks and
hawkish policies carry the day. A look at the upper hand that hawks had in
determining U.S. foreign policy as well as the electoral success that more
hawkish leaders seem to have enjoyed recently across the globe leads one to
believe that hawks and hawkish policies dominate in times of conflict.
Kahneman and Renshin~\cite{kahneman2007} suggest an explanation for a hawkish tendency
during conflict in terms of inherent human bias. They say that ``In terms of
potential conflict the same optimistic bias makes generals and politicians
receptive to advisers who offer highly favorable estimates of the outcome of
war." In analyzing war Baliga, Lucca and Sj\"{o}str\"{o}m~\cite{baliga2011} make a
similar point, they say that ``in a fully democratic country, a dovish bias
is replaced by a hawkish bias when the environment becomes more hostile.''
While accepting these explanations we want to examine whether hawkishness
can arise even in the absence of any optimistic bias (as in Kahneman and
Renshin) or any coordination failure (as in Baliga, Lucca and Sj\"{o}str\"{o}%
m). In the Discussion section we discuss the implications of our explanation
compared with some others in the literature. In particular, we examine
whether it could arise from the incentives of the political process itself.

If the political process itself is a cause of greater hawkishness, then one
may perhaps see why the then Lieutenant-General Moshe Yaalon, the Israeli
army's chief of staff (and no policy dove) infuriated Prime Minister Ariel
Sharon and Defence Minister Shaul Mofaz in 2003 by publicly questioning
Israel's tough policies in the West Bank and Gaza \cite{erlanger2005}.  Moreover, he criticized the then Prime Minister Olmerts ground
invasion of Lebanon as "It had no substantive security-political goal, only
a spin goal." \cite{mccarthy2006}. The case becomes even more interesting when we see how
the same Olmert adopted a dovish stance after losing the elections (and
effectively ending any electoral incentives he had) by saying ``We have to
reach an agreement with the Palestinians, the meaning of which is that in
practice we will withdraw from almost all the territories, if not all the
territories" \cite{mccarthy2008}. More recently, Benjamin Netanyahu's hawkish policy (whether strategic or
because he is an inherent hawk), has bought him electoral victory. Moving outside the Middle East conflict, India's incumbent Prime Minister Narendra
Modi has also been hawkish with neighboring Pakistan \cite{bali2019} and there is a view that it contributed to his huge electoral victory \cite{henue2019}. 

Indeed, politicians not known for their hawkish ideology may also be
strategically belligerent, then US President Barack Obama, when faced with a lowered
level of support among his electorate, gave a hawkish speech (ironically when
accepting the Nobel peace prize) signalling his intention to be tough in
fighting conflict \cite{southwood2009}.

Our paper seeks to understand how instances of hawkish behavior documented above arise in a political process. In order to formalize this idea, we embed a model of conflict (similar
to Grossman~\cite{grossman1994}) in a principal-agent model of political competition in
which the voters (the principal) use re-election as an incentive to induce
the incumbent political leader (the agent) to manage the conflict in an optimal way. The
conflict is modeled as dividing a fixed resource, for concreteness sake, a piece of land, between the citizenry and insurgents. The incumbent politician, as the representative of the citizenry chooses a policy, i.e., a division of land between the groups. Following the land division there may be insurgent activities (such as terrorism or other disruptive tactics) which impose a cost on the citizenry. Incumbents vary along two dimensions, their ability and ideology. Incumbent's ability denotes his success in
minimizing the cost to society from the insurgent activities, while ideology denotes the intrinsic utility from holding on
to land, which translates into his willingness to bear the costs of insurgency.

Voters hold incomplete information about the incumbent's type (ability and
ideology), and have to decide whether to re-elect the incumbent, taking
into account the mixed signal about ability and ideology that comes from
observing a particular policy chosen by the incumbent. In line with the illustrations of hawkish behavior discussed above, we wish to understand whether the incumbent has an incentive to choose hawkish or extremist policies. To make clear what we mean by this
extremism or hawkishness, a \textit{hawk} is a person who has a higher
utility (relative to the median voter) from retaining more land, while a \textit{hawkish policy} from a
particular leader is a policy more extreme than the one he would ideally
like to choose. We analyze if a hawkish policy emerges under the optimal
voting rule. We also analyze both the electoral success of politicians of
different ideologies under this voting rule and how the optimal voting
standard changes with changes in the rents from office as well as
ideology of politicians.

Our main finding is that electoral incentives lead politicians to choose policies that are more hawkish than those desired by the voter, conditional on the politician type. Furthermore, the political process, under certain conditions, confers a natural electoral advantage to incumbents which are intrinsically more hawkish than the voter. While we have a homogeneous population in our model, this result would hold with respect to the median voter's preferences in a heterogeneous population where voters have single peaked preferences over a single-issue policy space. 

{\color{black}{Thus, the electoral process leads to two kinds of hawkishness -- one,
politicians choose more hawkish policies (i.e., give less land) relative to what the voter would like them to implement, conditional on their ability (see Proposition 1); and two, politicians who intrinsically favor more
extreme policies are more likely to get elected (see Proposition 2).}} The intuition behind the first kind of hawkishness
is the following: politicians with higher ability are better able to successfully manage the higher costs emanating from more hawkish positions. As a result, they are able to signal their type
by choosing sufficiently extreme policies. Regarding the reason for the second kind of hawkishness note that, holding ability fixed, hawkish
politicians intrinsically prefer a more extreme policy as their 
utility from land is higher compared to more dovish ones. Hence, both high
ability and intrinsic hawkishness leads to tough policies that lead to
higher costs of conflict. If voters cannot distinguish between intrinsic
hawks (for many new leaders their ideology may not be fully known, at least
as regards national security and conflict negotiation) and high ability
types, leaders have to resort to behaving hawkishly to signal their ability
and inherent hawks have an intrinsic advantage in that. {\color{black}{We characterize the conditions under which the
equilibrium voting strategies turn out to be such that doves are replaced at
a faster rate than hawks.}}

{\color{black}{To present our results more precisely, we compare three outcomes: 1) the voter's ideal policy under complete information (i.e., the first best), 2) the voter's best outcome under incomplete information when he can ex-ante commit to a re-election strategy (i.e., the second best, which is also called the constrained Pareto optimum) and 3) the equilibrium (PBE) of the political process which is the outcome attained under incomplete information and without ex-ante commitment to a re-election strategy. Using the first best as the benchmark, we analyze the the re-election standards under the second best and the PBE for the case where incumbent's ideology is known to the voter but the ability is not.

In setting the re-election standard, with or without commitment, the voters face a trade-off. On the one hand, a higher re-election standard helps screen out low ability politicians. On the other hand, given that ideological hawks find it easier to meet the standard, when the incumbent is intrinsically more hawkish, screening for ability by raising the re-election standard may prove too costly. In this case voters may prefer not to raise the standard but instead lower it, accepting a poorer selection on ability in favor of a reduced cost of conflict. We show how this trade-off changes with the rents from office and incumbents ideology.

When the rents from office increase, politicians are more likely to want to meet the standard to
get re-elected and after a point voters may sacrifice selection to prevent the costs of conflict from being too high. In particular, this is true when politicians are more hawkish than the voter. In this case, both rents from office and inherent hawkishness incentivize the incumbent to meet the standard. In this scenario, voters may prefer to lower the standard and put up with low ability politicians getting re-elected to having too high a standard and facing very high costs of conflict. We show that the second best may indeed involve such non-monotonicity but the PBE does not.

With regards to the ideology, when the incumbent is a hawk, the second best re-election standard can be non-monotonic in ideology for the reasons outlined above. With dovish politicians, surprisingly, we find that that both the second best and the PBE may involve a hawkish policy. This occurs since the high ability doves face a trade-off between their implementing a policy more extreme than their ideal vs. a less hawkish policy being implemented by an average ability replacement. It can be that the expected cost of conflict is lower in the former case than the latter. 

The way the optimal standard varies with rents and ideology also allows us to answer another question: does the political equilibrium of this model achieve constrained Pareto optimality (i.e., the second best). We show that in
general it does not; the second best may, for instance involve re-electing all politicians which keeps the cost of conflict low though doing so is not incentive compatible and therefore not a political equilibrium. 

We then show that the incentive compatible standard is monotonic for the range we consider and,
in general, does not coincide with the second best. Thus, we show that there is a political failure,
i.e., the political process need not achieve the constrained Pareto optimum. If voters 
could bind themselves to re-electing every politician regardless
of their ability ex ante voter welfare may be higher. This implies that
tying the hands of the politician by making him agree to settlements
proposed by a neutral third party can be welfare improving. Thus, there is a
role for bodies like the UN to achieve Pareto improving settlements even in
a situation of conflict.}}

In sum, our model provides a new explanation of conflict, showing that it arises from re-election motives and that the incentive compatible reelection standard does not achieve the second best. This is a specific application of the notion of political failure. While such political failure and ‘excessive behavior’ has been analyzed in other contexts e.g. deficit financing, inefficient transfer to special interest groups (see the next section for details), this is novel as an explanation of conflict. Moreover, we show that the excessive behavior is always in one direction (i.e. induces hawkishness). Further, unlike other related models in the literature, we allow for variations in two dimensions-ideology and ability and show that the hawkish behavior results still hold and that the electoral process favors hawks.

\subsection*{Literature}
Our work is related to principal agent models starting from Spence~\cite{spence1973}. More specifically
our paper has modeling similarities with the so called incumbent-challenger
models which are essentially agency models of political competition. There
are a wide variety of incumbent challenger models in the literature
(see Besley~\cite{besley2006} for a good discussion of the literature), with the
simplest involving pure selection strategies to weed out bad politicians.
There are several well known models with electoral accountability
when politicians make unobservable choices~\cite{barro1973},~\cite{ferejohn1986},~\cite{rogoff1988},~\cite{rogoff1990} and~\cite{persson1997}. Rogoff and Sibert~\cite{rogoff1988} and Rogoff~\cite{rogoff1990}use this framework to
study political budget cycles, while the other papers look at pure moral
hazard problems. There are models with elements of both moral
hazard and adverse selection, in which politicians differ in both ability
(leading to adverse selection) as well as some kind of unobservability of
action (leading to moral hazard)~\cite{banks1993}. Coate and Morris~\cite{coate1995} consider the issue
of the form of transfers to special interest groups in an incumbent
challenger framework. They address a pure efficiency question: namely, given
that politicians may owe allegiance to special interest groups what is the
most efficient way to make transfers to such groups and is this efficient
form of transfer employed? In their framework, politicians do not
essentially differ in ability but in their preference over the transfers
they would like to make to special interest groups, in principle both
politicians could have chosen the best outcome. Such models of electoral
accountability have a fair amount of empirical support as well~\cite{besley1995}. In the context of conflict, Carter~\cite{carter2017} finds that electoral incentives can lead to conflict, in
particular they find term limited dovish leaders are less likely to initiate
conflicts, on average, compared to those who are electorally accountable while
there is no such relationship for hawkish leaders, consistent with our model.

Several papers have used the political agency model to shed light on
different forms of political failures arising from electoral concerns and
asymmetric information. Majumdar and Mukand~\cite{majumdar2004} consider a model of
political agency where leaders ability is modeled as his ability to pick a
socially desirable policy reform. The leader gets an interim signal as to
whether or not his reform is likely to succeed. While it is socially
desirable that a failing reform be scrapped, such an action will lead the
voters to realize that the politician took a wrong decision and therefore to
lower their estimate of his ability. Under certain conditions, this leads to
the politician resorting to a gamble by sticking to a failing policy. Aidt and Dutta~\cite{aidt2007} examine political failure arising out of the interaction
between observation lags, economic growth and a binding revenue constraint.
The political failure in their paper does not arise from asymmetric
information about politician ability (their politicians are homogenous) and
their aim is to look at the mix of short term vs. long term public good that
is provided because of these interactions. Unlike in our model, their policy
myopia is constrained optimum. In our analysis of second best we find a
result similar to that of Haan and Onderstal~\cite{haan2008} that randomization
over which politician to re-elect gives higher voter welfare than trying to
separate types (and randomization always dominates welfare under a pooling
equilibrium).

Our work is also related to rational choice explanations of terrorism and
the high cycle of violence which is based on terrorism as a strategic choice [~\cite{pape2003},~\cite{berman2003}]. Leadership in conflict is analyzed in a
complete information framework by Hess and Orphanides~\cite{hess1995} (see the Discussion section for a discussion of their paper) and is related to the literature that looks at whether multilateralism can provide an efficient
level of security and the structure it takes [~\cite{gupta2012},~\cite{gupta2010},~\cite{gupta2014}]. Schultz~\cite{schultz2003} analyses the
behavior of hawks and doves over the period that the US-USSR conflict
continued using a different definition of hawks and doves-he assumes that
doves are people who inherently have optimistic priors over the opponents
motives and the opposite for hawks. Some of the empirical evidence on the
policy positions of hawks and doves in that period is interesting and worth
analyzing to see how well it fits the predictions of our model. Broadly,
from our model, we expect to see hawks continue to support hawkish positions
while doves will fluctuate depending on their ability. This seems consistent
with the data presented in Schultz. None of these papers are however
concerned with the political failure arising out of asymmetric information.

There is a well established literature on conflict and the distribution of
resources across groups [~\cite{grossman1994}, \cite{conley2001}, \cite{hirschleifer1995}, \cite{garfinkel2007}, \cite{sanchez2009}]. We contribute to the literature
by analyzing what kind of outcomes arise when conflict resolution occurs via
the political process.

The rest of the paper proceeds as follows. In the Materials and Methods section we set up the model, and characterize the first best, the second best and the Perfect
Bayesian equilibrium (PBE) outcomes. The Results section compares the
second best and the PBE with each other and with the first best. This section provides our main insights about how the the political process leads to a  hawkish drift. The Discussion section considers some extensions and concludes.

*{Materials and Methods}
\subsection*{The Model}\label{model section}
We develop a simple two period model that formalizes the basic insight
discussed above. {\color{black}{The framework used here was developed by Grossman~\cite{grossman1994} and the specifics of the model are based on Bandyopadhyay and Oak~\cite{bandyopadhyay_oak_}.}} Let us denote by $Y \in \mathcal{R}_{+}$ the total amount of land, which is the
dimension of conflict between two groups: citizens denoted by $C$, and an
insurgent group (or protesters) denoted by $I$. While we use the nomenclature of an intra-country conflict for convenience, the model can capture any dispute over a fixed resource, including border disputes between two countries. The division of land is determined by an incumbent politician, specifically, the incumbent chooses a division $(y,Y-y)$ on behalf of group $C$ where $y$ is the
amount of land retained by the group $C$, and consequently, $(Y-y)$ is the land conceded to $I$. Hereafter, we will refer to the representative member of $C$ as \textit{the voter}, and the politician in office who decides the land division as \textit{the incumbent}. We can think of the representative member of $C$ as the median voter, an interpretation that holds under single peaked preferences. 

The voters preferences
over land division are represented by a utility function $u:[0,Y] \rightarrow \mathbb{R},u^{\prime }>0,u^{\prime \prime } \leq 0.$ The incumbents preferences over land division may differ from those of the voter. We denote the incumbents preferences by $\alpha \cdot u(y)$ where $\alpha \in (0, \infty)$ is a parameter that captures politicians ideology---a value of $\alpha < (>) 1$ means the incumbent is dovish (hawkish) relative to the voter. After observing the choice of $y$ made by the incumbent, the voter decides whether or not to re-elect him. 

Following the choice of $y$, some members of
group $I$ may undertake terrorist activities or violent insurgency. The cost imposed by this insurgency upon the voter is realized in the \textit{next} period and it is increasing in $y$. The politician in office in the next
period manages the cost of this conflict, which (given $y$) depends on his
ability $\theta$. Let $c(y,\theta )$ denote the expected cost of terrorism to the voter when the settlement is $y$ and the ability of the next periods politician is $\theta .$
We assume that $c_{1},c_{11}>0$ and $c_{2}<0.$ {\color{black}{In Appendix \textbf{I.A} we present a simple model that provides a micro foundation for the cost of insurgency.}} 

\noindent
\textbf{Sequence of actions}

\noindent
The game played is as follows:

\begin{itemize}
\item $t=0$: Nature chooses the type of incumbent $(\alpha, \theta)$. Voters (i.e. group $C$) set a re-election standard on $\hat{y}$ such that the incumbent is re-elected only if $y \geq \hat{y}$. 

\item $t=1$: The incumbent chooses policy $y\in \lbrack 0,Y]$. The voter votes to either re-elect or replace the incumbent. If the incumbent is voted out, he is replaced by another politician drawn from distribution $F(\theta ),$ with pdf given by $f(\theta
)$.

\item $t=2$: Payoffs are realized. 
 \end{itemize}
The expected payoff of the voter is $u(y)-c(y,\theta )$ if the incumbent is retained, and $u(y)- c(y,\theta')$ if the incumbent is replaced by a politician of ability $\theta'$. {\color{black}{The expected payoff of the incumbent differs from the voter in two ways. First, the intrinsic utility from retaining $y$ may differ from the voter by a multiple $\alpha$, and second, incumbent, if re-elected, gets additional payoff $r>0$, which may be interpreted as the rent from
holding office in the next period.}}  {\color{black}{Thus, the politicians payoff is}} $\alpha \cdot u(y) - c(y,\theta) +r$ if the incumbent is re-elected, and $\alpha \cdot u(y) - c(y,\theta')$ if he is replaced by a politician of ability $\theta'$. This payoff reflects the fact that the politician holding office in period 2 inherits the policy set by the incumbent and manages its cost to the society. In particular, we assume that the policy is not such that it can be instantly reversed. We do not use time subscripts as there is no discounting. Total payoffs are simply a sum of payoffs from both periods.

Throughout the paper we will use a simple formulation that allows us to
explicit numerical solutions and enables us to compare payoffs.
Specifically, we assume that 
\begin{equation}
u(y)\equiv y
\end{equation}%
and 
\begin{equation}
c(y,\theta )\equiv \frac{y^{2}}{2(1+\theta )}.
\end{equation}%
This is the simplest
formulation that allows us to convey the main results of the model. We further assume that the politician ability $\theta$ is drawn from a uniform distribution $[0,1]$ and is independent of the ideology parameter $\alpha$. The
qualitative features of the model do not depend on these functional form but
they enable us to numerically compare the first best, the second best and the PBE.  
{\color{black}{
\subsection*{Benchmark $\mathbf{(\alpha, \theta)}$-contingent contract}

Suppose that the incumbent politician's ideology as well as ability is known to the voter; we will denote an incumbents type as($\alpha,\theta$). The first best outcome (FB for short) sets an ideal benchmark for the highest utility attained by the voter if he could offer a type-contingent contract, subject to the incumbent's and voter's participation constraints. In order to define and analyze the first best, it is useful to define a few terms. If an incumbent knew/anticipated that he is not going to be elected, his optimal strategy would be to choose $y$ to maximize his expected utility given by
$$\alpha y - E_{\theta'} c(y,\theta') $$ where the expectation is taken over the replacement's ability $\theta'$, which yields $y$ that depends on $\alpha$ only, and which we denote by 
$$y_0(\alpha) \equiv \frac{\alpha}{\ln 2}.$$
Let the expected utility from replacing an incumbent (and therefore him choosing policy $y_0(\alpha))$ be denoted by $v_0(\alpha)$ for the incumbent, and by $V_0(\alpha)$ for the voter. It can be verified that,
$$v_0(\alpha) \equiv \frac{\alpha^2}{2\ln 2} \text{\hspace{1cm} and\hspace{1cm} } V_0(\alpha) \equiv \frac{\alpha}{\ln 2} \left(1 - \frac{\alpha}{2}\right).$$
The first best, can be formally defined as follows:
\begin{definition} 
	The first best contract for a type ($\alpha, \theta$) is a pair ($d^\ast, y^\ast$) whereby the agent implements policy $y^\ast \in Y$ and the voters re-election decision is $d^\ast \in \{\text{re-elect, replace}\}$ such that
	\begin{itemize}
		\item $d^\ast =$ re-elect, and  
		\begin{eqnarray}
			&y^\ast = \underset{y \in [0,Y]}{\arg\max} \hspace{1cm} &u(y) - c(y,\theta) \label{FB Problem}\\
			&\text{s.t.  } &\alpha u(y) -c(y,\theta)+r \geq v_0(\alpha) \nonumber
		\end{eqnarray}
	
		if 
		\begin{enumerate}
			\item a solution to the above problem exists, and
			\item $ u(y^\ast) - c(y^\ast, \theta) \geq V_0(\alpha);$
		\end{enumerate} 
		\item otherwise, $d^\ast =$ replace, and $y^\ast = y_0(\alpha).$
	\end{itemize}
When describing the first best for each possible type $(\alpha, \theta)$, we will use the notation $d^\ast(\alpha,\theta)$ and $y^\ast(\alpha,\theta)$.
\end{definition}
A complete characterization of the first-best is tedious as it depends on the different configuration of parameters $\alpha, \theta$ and $r$. We will characterize the first best in specific cases of interest in the later sections. However, some cases are worth mentioning here. 

If Problem (\ref{FB Problem}) has an interior solution, then it is $y(\alpha,\theta) = 1+\theta$, yielding the voter indirect utility $\frac{1+\theta}{2}.$ The voter will re-elect (replace) the incumbent if $\frac{1+\theta}{2} \geq (<) V_0(\alpha)$, i.e.
$$\theta \geq (<) \frac{\alpha(2-\alpha)}{\ln 2} -1 .$$
Thus, we have the following lemma. 
\begin{lemma}
	If $\alpha, \theta$ and $r$ are such that Problem (\ref{FB Problem}) has an interior solution, then the first best policy is given by,
	\begin{eqnarray}
		y^\ast(\alpha,\theta) = 1+\theta,\hspace{0.25cm} &d^\ast(\alpha,\theta) = \text{re-elect} \hspace{0.5cm}&\text{if}\hspace{0.5cm} \theta \geq \frac{\alpha(2-\alpha)}{\ln 2} -1, \\
		y^\ast(\alpha,\theta) = \frac{\alpha}{\ln 2},\hspace{0.25cm} &d^\ast(\alpha,\theta) = \text{replace} \hspace{0.5cm}&\text{if}\hspace{0.5cm} \theta < \frac{\alpha(2-\alpha)}{\ln 2} -1.	
	\end{eqnarray}	
\end{lemma} 
Problem (\ref{FB Problem}) is more likely to have an interior solution if $i)$ $r$ and $\theta$ are large and $ii)$ if $\alpha$ is not too large or small relative to 1. 

In the following sub-sections we analyze  a more realistic setting where the voter cannot observe the incumbents ability. In such a setting, the voters re-election decision can be contingent only on the observed policy choice. The second best refers to the case where the voter can credibly commit to a re-election standard. The Perfect Bayesian Equilibrium (PBE) refers to the case where no such commitment is possible.}}

\subsection*{The Second Best, or Equilibrium under Commitment}
Now suppose that the voters cannot observe the incumbents type but can
commit to a re-election strategy based on his choice of $y$. In this section we will assume that the politicians ideology ($\alpha$) is known to the voter. We will discuss the implications of relaxing this assumption in the next section. We will 
assume that voters use a cut-off strategy: one where they re-elect the
incumbent if and only if $y\geq z$. We restrict attention to cut-off strategies to aid comparison with the PBE in monotone beliefs which involves cut-off strategies. {\color{black}{This restriction,which is standard in the literature~\cite{coate1995}, is the equivalent of having monotone beliefs and rules out PBE supported by unappealing off the path equilibrium beliefs. Throughout when we refer to second best and PBE, we restrict attention to cutoff strategies.}}

{\color{black}{We now look at the incumbent's strategy in response to a re-election standard $z$ set by the voter. The incumbent will choose to fulfill the standard $z$ and get re-elected, if there exists a $y \geq z$, such that}}
\begin{equation}
	\alpha y-\frac{y^{2}}{2(1+\theta)} +r \geq \alpha y -E_{\theta'}c(y,\theta').
\end{equation}
Holding ideology fixed, the behavior of politicians of different ability levels can be described as follows: 
\begin{itemize}
	\item Incumbents with "low ability" will prefer to drop out of the race than meet the re-election standard, i.e. $\forall \theta \leq \theta_1$
	$$\alpha {z}-\dfrac{{z}^{2}}{2(1+\theta _{1})}+r  \leq \alpha
	y_{0}(\alpha) -\int_{0}^{1}\dfrac{y_{0}(\alpha)^{2}}{2(1+\theta )}d{\theta };$$
	
	\item Incumbents with "intermediate ability" will just meet the re-election standard, i.e., $\forall \theta \in (\theta_1, \theta_2]$ we have 
	$$\alpha {z}-\dfrac{{z}^{2}}{2(1+\theta _{1})}+r  > \alpha
	y_{0}(\alpha)-\int_{0}^{1}\dfrac{y_{0}(\alpha)^{2}}{2(1+\theta )}d{\theta }$$ and $z \geq \alpha(1+\theta)$ -- the optimal policy of the incumbent of type $(\alpha,\theta)$ if he is guaranteed re-election;
	
	\item Incumbents with "high ability" $\theta > \theta_2$ implement their optimal policy subject to guaranteed re-election, ie, they choose $\alpha(1+\theta)$.
\end{itemize}

We first compute $\theta _{1}$. This is the lower limit of the ability of
the politician who meets the standard. Evidently, at this limit, the
politician is indifferent to re-election and being replaced, hence

\begin{eqnarray}
	\alpha {z}-\dfrac{{z}^{2}}{2(1+\theta _{1})}+r &=&\alpha
	y_{0}-\int_{0}^{1}\dfrac{y_{0}^{2}}{2(1+\theta )}d{\theta }  \notag \\
	\alpha {z}-\dfrac{{z}^{2}}{2(1+\theta _{1})}+r &=&%
	\dfrac{\alpha ^{2}}{2\log 2}  \notag \\
	\theta _{1}(z) &=&\dfrac{{z}^{2} \cdot \ln 2} {(\alpha z%
		+r) \cdot 2 \ln 2 -\alpha ^{2}}  -1  \label{thetaone}
\end{eqnarray}

We now compute $\theta _{2}$ which is the upper limit of the re-election
standard. $\theta _{2}$ is the value of $\theta $ at which
the politicians ideal policy coincides with the standard ${z}$:

\begin{eqnarray}
	z &=&\alpha (1+\theta _{2})  \notag \\
	\theta _{2}(z) &=&\dfrac{z}{\alpha }-1   \label{thetatwo}
\end{eqnarray}

Given the expressions for $\theta_1(z)$ and $\theta_2(z)$, the expected utility of the voter can be written as
\begin{eqnarray}
	W(z) &=& \displaystyle \bigg[\displaystyle\int_{\theta _{1}(z)}^{\theta
		_{2}(z)}\bigg({z}-\dfrac{{z}^{2}}{2(1+\theta )}\bigg)d{\theta }+%
	\displaystyle\int_{\theta _{2}(z)}^{1}\bigg(\alpha(1+\theta)-\dfrac{(\alpha(1+\theta))^{2}}{2(1+\theta )}\bigg)d%
	{\theta } \notag \\
	&+& \displaystyle\int_{0}^{\theta _{1}(z)}\bigg({y_{0}}-\dfrac{{y_{0}}^{2}%
	}{2}\ln 2\bigg)d{\theta }\bigg]  \notag \\
	&=& \displaystyle \bigg[z(\theta _{2}(z)-\theta _{1}(z))-\frac{%
		z^{2}}{2} \cdot \ln \bigg(\dfrac{1+\theta _{2}(z)}{1+\theta _{1}(z)}\bigg)+\bigg(%
	\alpha -\frac{\alpha ^{2}}{2}\bigg)\bigg(\frac{3}{2} \notag \\
	&+& \dfrac{\theta _{1}(z)}{%
		\ln 2}-\theta _{2}(z)-\frac{\theta _{2}(z)^{2}}{2}\bigg)\bigg]  \label{W}
\end{eqnarray}

{\color{black}{The above expression has three components which correspond to the three possible types of choices made by the politicians, depending on their abilities, given the re-election standard. Note that the cut-off strategy of the voter include, as special cases, setting $z$ equal to 0 (always re-elect) or infinity (always replace). The second best re-election standard is $\bar{y} \equiv \arg\ max W(z)$.}}

If the voter chooses to always replace the incumbent, then the
expected utility of the voter is given by $V_{0}=\frac{\alpha }{\ln 2}(1-\frac{\alpha}{2}).$ If
the voter were to always re-elect the incumbent, the expressions differ depending on whether the rents
from office $r$ are high enough for voluntary dropout never to occur. As our
purpose is to look at a world where the rents from office cause a
divergence between the voter and the politicians behavior, one particular case of interest is when the 
rents from office are high enough such that no one voluntarily drops out.
In that case, the utility is given by the expected utility of voter $V_{E}=\displaystyle%
\left(\alpha -\frac{\alpha ^{2}}{2}\right)\int_{0}^{1}(1+\theta )d\theta
=\left(\alpha -\frac{\alpha ^{2}}{2}\right)\frac{3}{2}=\frac{3\alpha }{2}%
\left(1-\frac{\alpha }{2}\right).$ In this formulation, the utility from
re-electing everyone is again bigger than replacing everyone as $V_{0}<V_{E}.$

\subsection*{Perfect Bayesian Equilibrium or Equilibrium Standard without Commitment}

The second best formulation above assumes that there is no additional
constraint in terms of incentive compatibility of the voting standard. That
is, the voters cannot renege from the re-election strategy after they
observe the politicians choice of $y$. However, for the re-election standard to
be incentive compatible, given the beliefs of the voter about the expected
ability of the incumbent, his voting decision must be optimal. We
will use the Perfect Bayesian Equilibrium (PBE) as the solution concept to analyze the outcome of a game in which the
voter is unable to commit to a  re-election standard. The
PBE is formally defined as follows:
\begin{definition}
A PBE of the game is (1) a \ re-election standard
set by the voters $\widehat{y}$; (2) a choice of $y(\theta ,\widehat{y})$,
by the incumbent politician, given the re-election standard $\widehat{y}$;
(3) a set of beliefs $b:\left[ 0,1\right] \rightarrow B(\theta )$ i.e. a
mapping from the $y$ chosen into a probability distribution over types
denoted by $B(\theta )$ and (4) a re-election decision upon observing the
policy choice $y$, such that

1) the politician of type $\theta $ chooses $y$ to maximize his expected
payoff, which is $\frac{\alpha^2}{2 \ln 2}$ if $y<\widehat{y}$, and $\alpha u(y)-c(y, \theta )+r$ if $%
y\geq \widehat{y};$

2) the voters choose $\widehat{y}$ to maximize their expected utility given
the politicians best response $y(\theta ,\widehat{y});$

3) voters beliefs about the incumbents type are formed using the Bayes Rule
(whenever possible); 

4) Voters re-elect a politician only if given their beliefs (as in 2) their expected
utility from re-electing is greater than from replacement; and

5) Voters re-election strategy (as in 3) is consistent with the re-election standard (as in 1).
\end{definition}

{\color{black}{Additionally, we restrict attention to PBE in monotone beliefs, i.e., the use of cutoff strategies.}} In particular, the voters expected utility from retaining those politicians
who just meet the standard must be bigger than the (constant) expected
utility from replacement, i.e., $V_{0}$. This gives us the following
incentive compatibility constraint i.e. 
\begin{equation} \label{IC}
	\int_{\theta _{1}(\hat{y})}^{\theta _{2}(\hat{y})}\left[ \hat{y}-\frac{\hat{y%
		}^{2}}{2(1+\theta )}\right] \frac{d\theta}{\theta _{2}-\theta _{1}} \geq 
	\frac{\alpha}{2\ln 2}(1- \frac{\alpha}{2}) \left( \equiv V_0(\alpha) \right)
\end{equation}%
Thus, the equilibrium standard without commitment, $\hat{y}$, maximizes the expression for the expected utility function (as in the previous subsection):
\begin{equation}y(\theta _{2}(y)-\theta _{1}(y))-\frac{y^{2}}{2} \ln \bigg(\frac{1+\theta _{2}(y)}{1+\theta _{1}(y)}\bigg)+\bigg(\alpha -\frac{\alpha ^{2}}{2}\bigg)\bigg(\frac{3}{2}+\frac{\theta _{1}(y)}{
	\log 2}-\theta _{2}(y)-\frac{\theta _{2}(y)^{2}}{2}\bigg)
\end{equation}
but subject to the incentive compatibility constraint in equation (\ref{IC} as given above. This constraint requires that
the standard must be sufficiently high such that some ability types find it
costly to mimic it in order to seek re-election. The chosen re-election standard under PBE is given by $\hat{y}$.

{\color{black}{Note that we cannot find closed form solutions for the above equations in the general range of parameters. We solve the maximization problem using Mathematica as well as R to see how $y$ varies across the relevant range of $\alpha$ and $r$.}}

It is worth noting that the PBE may involve
replacing everyone as it satisfies the incentive compatibility constraint, however it should be
clear that re-electing everyone does not satisfy the additional constraint. This is because, if incumbents chose their optimal strategy when guaranteed re-election (i.e. $y = \alpha(1+\theta)$), it perfectly reveals their type and for incumbent with $\theta < \frac{1}{\ln 2}$ the voters would be better off, ex-post, by reneging on their promise to re-elect the incumbent. Thus, in general, the second best and the equilibrium standard without
commitment will not coincide. We illustrate this in more detail in the next
section.

\section*{Results}\label{analysis section}
In this section we will analyze the policy outcomes under the Second Best and the PBE as defined in the previous section. In both cases, the voters seek to screen high ability incumbents by setting a re-election standard such that the expected ability of an incumbent who meets the standard is (weakly) greater than those who do not. Under the PBE it is further required that the standard should be ex-post credible, that is the expected utility of re-electing an incumbent meeting the re-election standard is greater than that of the expected replacement.

We are particularly interested in understanding when policy choice of an incumbent is more (or less) hawkish compared to the first best policy as defined in the previous section and (partially) characterized in Lemma 1. 
This is a useful benchmark against which we can compare the Second Best and PBE policy choices of incumbents. 
{\color{black}{\begin{definition}
	An outcome, under the second best or PBE, is said to exhibit a hawkish drift if each incumbent who gets re-elected chooses a policy $y \geq y^\ast(\alpha,\theta)$, his first best policy, with a strict inequality for non-zero measure of incumbents.
\end{definition}}}

There are two reasons why the second best and the PBE outcomes would not coincide with the first best. First, due to the presence of the rents from office ($r$), incumbents have an incentive to meet the re-election standard even if it is higher than their ideal policy. This results in some low ability incumbents, with $\theta < \frac{1}{\ln 2}$, meeting the standard and therefore getting re-elected. Voters will anticipate this response from the incumbents in setting the re-election standard, which may therefore be more hawkish than the first best in order to screen out low ability incumbents. Second, when an incumbents intrinsic preference over $y$ differs from the voters, voters may use re-election standard to incentivize the incumbent into choosing a policy more aligned with that of the voters. Such standard may induce a hawkish (or dovish) policy choice by the incumbent.

As a very special case, when $\alpha =1$ and $r =0$, the voters and politicians have perfectly congruent preferences. That being the case, the second best re-election standard ($\bar{y})$ and the PBE standard $\hat{y}$ both equal $1+ \frac{1}{\ln 2}$ which leads to incumbents above (below) $\theta = \frac{1}{\ln 2}$ getting re-elected (replaced) and therefore the policy outcome coincides with the first best. 

\subsection*{Homogeneous Case}
To fix ideas, lets first focus on the case where the ideology of the incumbent is same as that of the voter, i.e., $\alpha = 1$. Plugging $\alpha =1$ in equations \ref{thetaone} and \ref{thetatwo} we get 

\begin{equation}
	\theta _{1}(\bar{y})\equiv \frac{\bar{y}^{2}}{2\left(\bar{y}+r-\frac{1}{2\ln
			2}\right)}-1 \nonumber
\end{equation}
and 
\begin{equation}
	\theta_2(\bar{y}) = \bar{y} -1 \nonumber
\end{equation}
Substituting these in equation \ref{W}, we get the expected utility of the voter as
$$W(\bar{y}) = \displaystyle \int_{0}^{\theta _{1}(\bar{y})} \frac{1}{2\ln 2} d\theta
+ \displaystyle \int_{\theta _{1}(\bar{y})}^{\theta_{2}(\bar{y})}[\bar{y}-\frac{\bar{y}^{2}}{2(1+\theta )}] d\theta
+ \displaystyle \int_{\theta _{2}(\bar{y})}^{1}\frac{1+\theta }{2}d\theta .  $$

Voters will choose optimal re-election standard to maximize the above function. Denote by $\bar{y}^{\ast }\equiv \underset{\bar{y}}{\arg \max }\ [W(\bar{y})]$
the second-best re-election standard, i.e., the standard chosen by the voter if he can
commit re-electing an incumbent who meets, or exceeds, the standard.
It is not necessary for the second best standard
to be a strictly positive finite number. It may well be the case that either of two polar strategies viz.
always re-elect ($\bar{y} =0$) or always replace ($\bar{y} = \infty$)
is the optimal choice. Under the always replace standard, the expected utility of the
voter is given by $V_{0}=\frac{1}{2\ln 2}$, while under always re-elect, the
expected utility of the voter is
given by $\displaystyle\int_{0}^{1}\left(\frac{1+\theta }{2}\right)\:d\theta
=\frac{1}{2}{\left(\theta +\frac{\theta ^{2}}{2}\right)|}_{0}^{1}=0.75$. In
this formulation, the utility from re-electing everyone is bigger than
replacing everyone as $0.75>\frac{1}{2\ln 2}\approx 0.72135.$ This gives rise to the following remark:
\begin{remark}
	The second best policy is either to always re-elect the incumbent, or to use
	a cut-off strategy $\bar{y} \in (0, \infty)$. The latter is the optimal choice only if $V(%
	\bar{y})\geq 0.75$.
\end{remark}
It follows that the Welfare of the voter under second best is
bounded below by 0.75 since the strategy of always re-elect is a feasible
strategy for the voters, and it yields payoff 0.75.

As discussed in the previous section, the PBE, i.e. the equilibrium without commitment puts an
additional constraint on the re-election standard in terms of the incentive
compatibility constraint \\i.e. $\displaystyle\int_{\theta_1(\bar{y})}^{\theta_2 (\bar{y})}%
\left[ \bar{y}-\frac{\bar{y}^{2}}{2(1+\theta )}\right] d\theta \geq \frac{1}{%
	2\ln 2},$ when $r$ is high enough, this holds with equality and everyone
gets replaced yielding a welfare of $\frac{1}{2\ln 2}.$ As noted, always re-elect is not incentive compatible, and hence cannot be part of a PBE.

We now provide a numerical example of the first best, second best and the PBE for specific numerical values. {\color{black}{In \nameref{Appendix II} we provide a more detailed set of calculations for a more general set of parameters. 

\begin{example}
	Consider the case $\alpha = 1, r= 0.2$. Under the second best re-election standard is $\bar{y} = 1.691$, which yields $\theta_1 = 0.222$ and $\theta_2 = 0.691$. That is, incumbents with ability below $0.222$ choose $y_0 = \frac{1}{\ln 2} = 1.443 < \bar{y}$ and therefore do not get re-elected, those with ability in the range $[0.222,0.691]$ choose to just meet the re-election standard $\bar{y} =1.691$, and those with ability $\theta >0.691$ choose $y = 1+\theta$. This outcome exhibits hawkish drift since the incumbents in the range $[0.222, 0.691)$ implement policy greater than their choice under the first best.
	
	Also note that the selection under the second best is significantly worse relative to the first best since, under the second best, those with ability greater than 0.222 get re-elected, whereas under re-election threshold under the first best is 0.443.
	It can also be verified that $\bar{y} = 1.691$ satisfies the incentive compatibility constraint given by equation \ref{IC} and hence the second best outcome is also the PBE.
\end{example}
However, the second best outcome need not always be attained under the equilibrium standard (i.e. the PBE), as seen in the following examples. 
\begin{example}
	Consider the case $\alpha = 1, r= 0.3$. In this case, the second best re-election standard is $\bar{y} =1.76$, leading to $\theta_1 = 0.157$ and $\theta_2 = 0.76$. However, the second best does not satisfy the incentive compatibility constraint. The PBE re-election standard is $\hat{y} = 1.77$ which yields $\theta_1 = 0.161$ and $\theta_2 = 0.77$. In this case, both the second best and the PBE exhibit a hawkish drift.
\end{example}
\begin{example}	
	More interestingly, consider the case $\alpha = 1, r=0.35$. In this case, the second best standard is to always re-elect, leading to $\theta_1 =\theta_2 = 0$. The PBE however, involves setting $\hat{y} = 1.8$ leading to $\theta_1= 0.134$ and $\theta_2 = 0.8$. In this case the PBE exhibits a hawkish drift.
\end{example}}}
We plot in the Fig 1 below, the second best re-election standard ($\bar{y}$) and the PBE standard ($\hat{y}$) against $r$. For reference we also plot the average policy implemented by incumbent in the range $[\theta_1, \theta_2]$ under the first best. This provides a natural metric for the extent of hawkish drift under the second best and the PBE.

%\centerline{\bf{Fig 1}}
\begin{figure}[H]
\includegraphics[height=8.0cm,width=12.0cm]{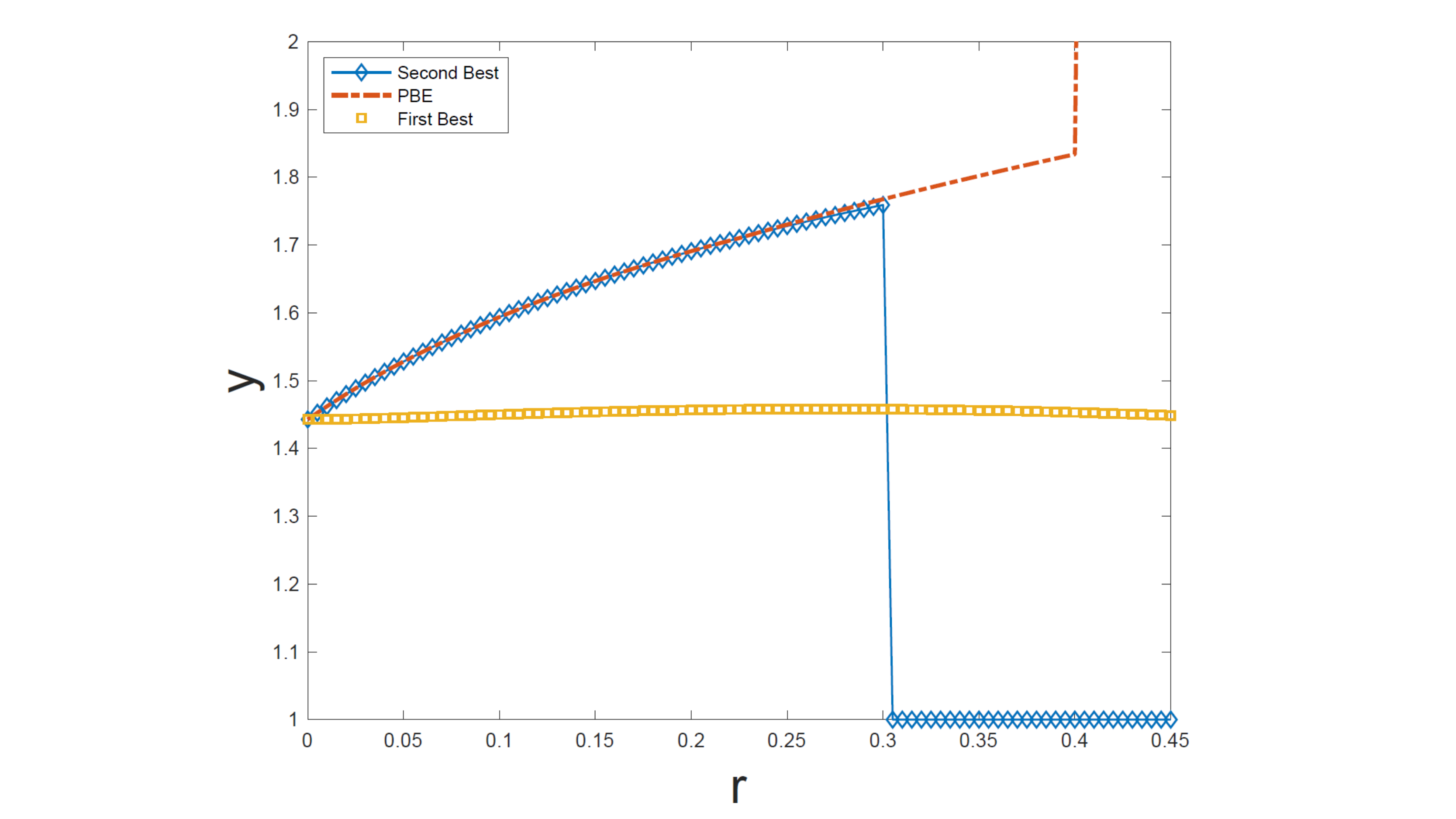}
\caption{Re-election Standards under Second Best and PBE, and FB with $\alpha =1$.}
\label{fig1}
\end{figure}

As we can see, when $r$ is small (approx less than 0.225), the re-election standard under the PBE is also second best. For higher values of $r$, however, the PBE is not second best as it violated the incentive compatibility. In fact, for very high values of $r$ (approx higher than 0.3) the second best involves always re-electing the incumbent.

{\color{black}{There are a few things worth noting. First, both the second best and the PBE standards are increasing in $r$. Second, both the Second Best and the PBE are typically higher than the First Best. This is one type of hawkish drift. Note the hawkish drift in the $\theta_1, \theta_2$ range. Both the Second Best and the PBE involve Hawkish drift (i.e., $\bar{y}^\ast, \hat{y}^\ast >y^\ast$). The average ability of the incumbent who gets reelected drops as $r$ goes up.}}

\noindent
The following proposition summarizes our results. 
\begin{proposition}
	The Perfect Bayesian equilibrium re-election standard is increasing $r$ and is weakly higher than the second best re-election standard. The two standards do not always coincide, in particular, there is a cutoff value of r at which they diverge. {\color{black}{Moreover, both the second best and the PBE involve hawkish drift as defined above.}}
\end{proposition}

\subsection*{General Case: How $\hat{y}$ Varies with Ideology and Rents}
{\color{black}{In the previous sub-section we saw how presence of rents from office ($r>0$) can lead to a hawkish drift under both the second best and the PBE. We now look at how the divergence of ideology ($\alpha \lessgtr 0)$ is a further source of hawkish drift, and the reasons thereof.
		
First consider the case when $\alpha > 1$, i.e., the politicians ideology is intrinsically more hawkish. Given any re-election standard and the value of $\alpha$, as $r$ increases, the average ability of the incumbents who meet the standard decreases. With $\alpha>1$, this leads to the voter facing a trade-off between screening, i.e., setting a higher $y$ to re-elect a higher ability incumbents, and a higher cost of conflict due to the policy being higher than their ideal policy given the politician ability. Depending on this trade-off, the voters will set a higher or lower re-election standard as $r$ increases. In the subsequent examples we will examine how the second best and the PBE  change, for a given $\alpha >1$, when $r$ increases.
\begin{example}
	Consider $\alpha = 1.1$, with $r \leq 0.08$. In this case, the first best policy, contingent on $\alpha = 1.1$, involves re-electing incumbents with ability greater than $\frac{1}{\ln 2} (\approx 0.443)$ and at this threshold, the policy is $\alpha *\frac{1}{\ln 2} \approx 1.587$. At $r=0$, the voters second best re-election standard is also $\bar{y} =1.587$, which is the ideal outcome for type $\alpha =1.1$, but this is higher than first best. This outcome is also the PBE.
	
	However, as $r$ increases, under the second-best, the voter may find it optimal to lower the standard. For instance, when $r=0.05$, the second best standard is $\bar{y} = 1.496$, leading to $\theta_1 = \theta_2 =0.360$. Thus the second best seeks to curb, to some extent, the intrinsic hawkishness of the low ability incumbents by committing to re-elect them at the standard lower than their first best. This is not a PBE, since the re-election strategy is not ex-post credible whenever an incumbents policy choice reveals his ability to be below $1.443$. 
	
	In this case, the PBE involves raising the re-election standard to the point where the expected quality of the incumbents meeting it makes the re-election strategy credible. The PBE standard when $r=0.05$ is $\hat{y} =1.665$. Under the PBE we have, $\theta_1 =0.374$ and $\theta_2 = 0.513$.
\end{example}
However, with a further increase in $r$, the benefit of curbing hawkishness by lowering the standard is outweighed by the cost in terms of poorer selection on ability. This leads the voter to raise the standard, as shown in the next example.
\begin{example}
	Consider $\alpha = 1.1$, with $r = 0.15$. In this case, the second best standard is $\bar{y} = 1.501$ which, while still below the first best of $1.587$, is higher than the second best standard in the previous example with $r=0.05$. Under this standard $\theta_1 = 0.214 $ and $\theta_2 = 0.369$. Since the cutoff values are both below the $0.443$ threshold, the second best is not a PBE.
	The PBE in this case is $\hat{y} = 1.778$, which induces $\theta_1 = 0.282$ and $\theta_2 =0.616$.
\end{example}
To summarize, the two examples demonstrate the trade-offs facing a voter. A higher standard improves selection but comes at the cost of a more hawkish policy. When there are no rents from office, the voter is unable to incentivize the politician to deviate from his optimal policy. As $r$ increases, voter can sacrifice selection to an extent by promising re-election, curbing low quality ‘hawks’. As $r$ increases further, more and more low quality politicians want to get re-elected (i.e. $\theta_1$ drops). In response the voter has to increase the standard again. This non monotonic feature of the second best when politicians are hawks shows the interplay between inducing better selection vs. facing higher costs of conflict.
This non-monotonicity is however not credible i.e. in the absence of commitment this cannot happen. Thus, the PBE is monotonic in $r$, inducing increases in the standard as r increases. 

We now consider the case of $\alpha <1$, i.e., the case where the incumbents ideology is dovish relative to the voter. In this case the voter does not face the trade-off he faced in the $\alpha>1$ case. The lower the $\alpha$ the easier it is for him to screen for a high quality by setting a high re-election standard. As the next series of examples shows, this generates a hawkish drift wherein the re-election standard is higher than not only the incumbents first best, but that of the voter as well.
\begin{example}
	Consider $\alpha = 0.9$. Suppose that $r=0$. In this case the first best policy involves selecting incumbents with $\theta = \frac{1}{\ln 2} \approx 0.443$ with the corresponding standard at this threshold to be $1.298$. The second best standard, however, is $\bar{y} = 1.559$ leading to $\theta_1 = 0.484$ and $\theta_2 = 0.732$. This outcome is supported under PBE as well, since the average quality of those meeting the standard exceeds $0.443$. 
\end{example}
The above example provides an interesting insight about hawkish drift. In the above example, the hawkish policy is used to drive selection and to correct for the inherent dovish bias of the incumbent. Even though the policy is too high for even some high ability incumbents first best they prefer to implement it rather than having their ideal policy implemented by an average quality replacement. Next we examine the effect of $r>0$ on the re-election standard. 
\begin{example}
	Consider $\alpha =0.9$ and $r= 0.1$. In this case the second best re-election standard is $\bar{y} =1.713$, leading to $\theta_1 = 0.388$ and $\theta_2 = 0.901$. This outcome is supported as PBE as well.
	
	As we can see, relative to the case with $r=0$, the selection is worse -- ($\theta_1 = 0.388$) vis-a-vis 0.484. This is due to fact that rents induce low ability politicians to meet the re-election standard. And the re-election standard is higher as compared to the $r=0$ case. Thus, voter balances the cost of selection against a higher cost of conflict by setting a standard that allows some below average doves to get re-elected.
\end{example}
}}
These examples help us understand how the second best and the PBE change as ideology ($\alpha $)
and rents from office ($r$) change. Recall the case where the politicians
ideology is the same as the voter. As noted earlier, there are two opposing
changes. As $r$ increases, the divergence between the politician and the
voters preferences increase. Thus, each ability type is more likely to meet
a given performance standard than not meet it. To be able to screen out bad
politicians, one has to choose a higher $\hat{y}$ (or $\bar{y}$, as the case may be, depending on whether we are looking at the PBE or the second best). However, this comes at
the cost of excessively high $\hat{y} (\bar{y})$ compared to the first best{. }As $r$
increases further, the cost of excessively hawkish policy may well outweigh
the gain from selection and $\hat{y} (\bar{y})$ may begin to dip. In that case we could even see the 
$\hat{y} (\bar{y})$ to reach a maximum and then fall. It would fall to either
selecting everyone or replacing everyone out. 

For our given utility
function, utility is always higher from re-electing everyone than replacing
everyone. Thus, the second best becomes a case of choosing between the
optimal $y$ found from maximizing the $W$ function and comparing with
re-electing everyone. Given the functional form we have chosen this is a
comparison between comparing the welfare from maximizing $W$ with 0.75.(
Recall welfare from re-electing everyone is $\frac{3\alpha }{2}\left( 1- \frac{\alpha }{2}\right) ,$ for $\alpha =1,$ this is 0.75.)

%\centerline{\bf{Fig 2}}
\begin{figure}[H]
\includegraphics[height=8.0cm,width=12.0cm]{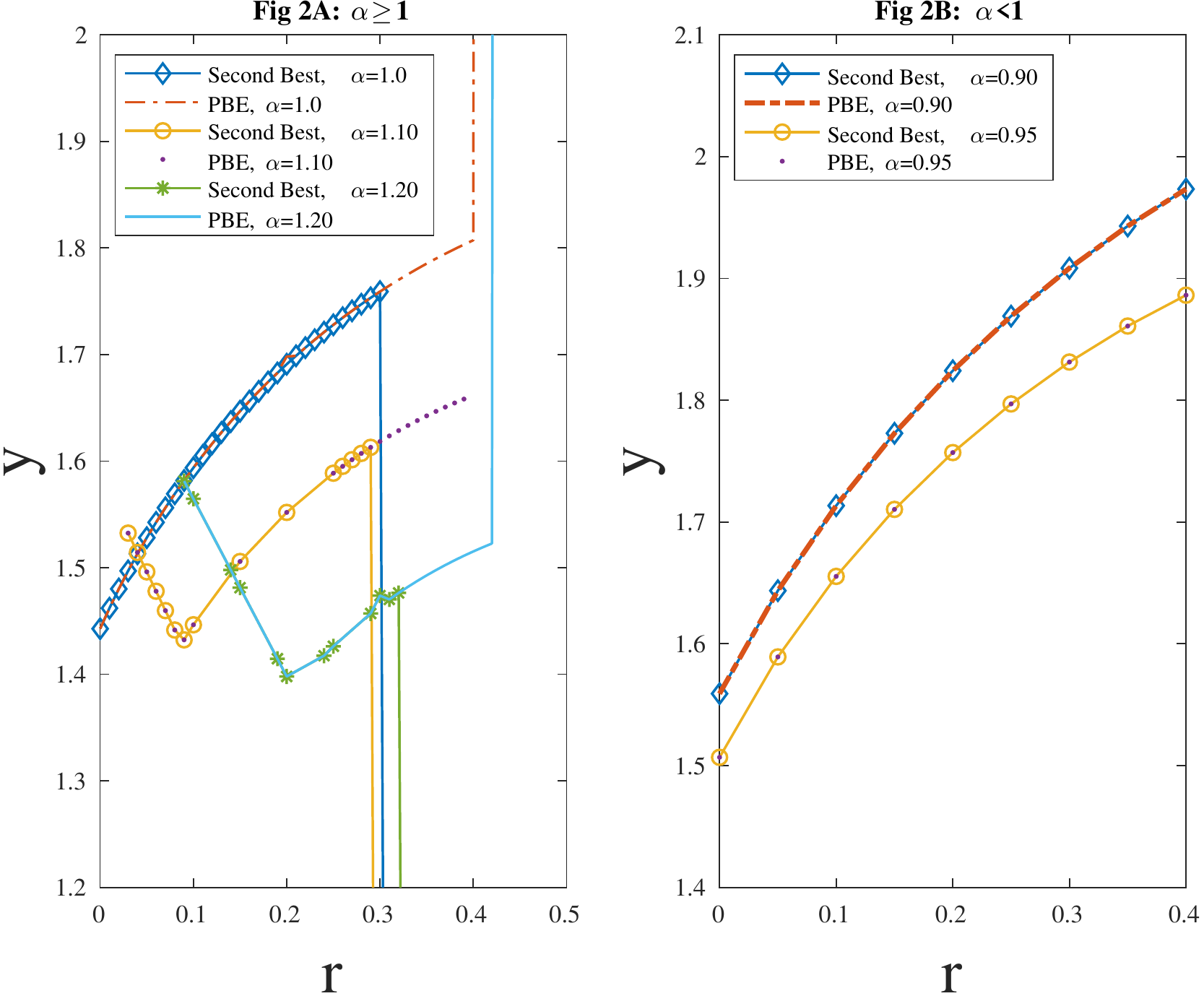}
\caption{Characterizing the re-election standards under Second Best and PBE: Fig 2A represents the case for
$\alpha \geq 1$; Fig 2B represents that for $\alpha < 1$.}	
\label{fig2}
\end{figure}

{\color{black}{As Fig 2A above shows, for $\alpha >1$, $\bar{y}$ is non-monotonic in $r$ and drops to 0 after around $r=0.3$ to satisfy the $V_{E}$ constraint. Fig 2B, on the other hand, represents a monotonic profile for $\alpha < 1$ without the discontinuities evident in Fig 2A. Fig 2 also illustrates how the second best and the PBE vary for both hawks and doves.}}

The characterization of how the voting standard changes when $\alpha$ is not equal to 1 is complex and the examples above provide the interesting trade-offs involved. In \nameref{Appendix II} we provide a table showing the change in $\bar{y}$ and $\hat{y}$, and the induced $\theta_1$ and $\theta_2$ for a range of values. Here, we summarize the essential features of the second best and the PBE.

Holding ability fixed, hawkish politicians find it easier to meet any standard. To get selection the voters needs a high $y$, but that may prove too costly in terms of welfare as it is beyond the voter’s ideal level of hawkishness even adjusting for quality. To strike the balance, $y$ may be optimally pushed down but that ‘lower’ $y$ may run into a credibility problem, i.e., the average of the people who bunch at the standard may be too low and are better replaced. Hence the second best is not a PBE when voters second best optimally causes $y$ to drop for a range as $r$ increases. 

{\color{black}{More strikingly, there is hawkish drift with politicians who are doves too. Even without any inducement from having $r>0$, doves who are above average in ability may not drop out if the policy is more hawkish than their first best. They face a trade-off, suffer a more hawkish policy than they would like but implement that (by meeting the standard) at a lower cost than their ideal policy implemented by their replacement whose expected quality is lower than their own. This leads to a hawkish drift again for a range if theta and this is reinforced as r increases. Both second best and the PBE are monotone in $r$ in this case.

We now turn to what happens when ideology as well as ability is unknown.}}

\subsection*{Model when both Ideology and Ability are Unknown}

We now consider the case where the voter cannot condition his re-election standard on ideology or ability. We then show that there is a possibility of a hawkish drift in this case as well. This result is intuitive: since we know that there exists a hawkish drift for a wide range of $\alpha$, the optimal policy when $\alpha$ is not known will involve setting a standard that it some convex combination of the $\alpha$-contingent re-election standards, and will therefore exhibit a hawkish drift. The following example is illustrative.
{\color{black}{
\begin{example}
	Suppose $r=0.15$ and the incumbent ability is either $\alpha =0.95$ or $1.05$, each with probability 0.5. The second best standard is $\bar{y} =1.65$. Given this standard, if $\alpha =0.95$, we have $(\theta_1,\theta_2) = (0.278, 0.741)$, and if $\alpha = 1.05$, we have $(\theta_1,\theta_2) = (0.253, 0.575)$. In either case, there is a hawkish drift for ability types $<0.65$. Since the PBE standard is weakly higher than second best, the PBE exhibits a hawkish drift as well.
\end{example}

To generalize from the above example, denote by $W(z,\alpha)$ the maximization problem as described in equation (\ref{W}) associated with ability type $\alpha$, and let $\bar{y}(\alpha)$ denote the corresponding second best standard. Suppose $\alpha$ is unknown and has probability distribution $F(\alpha)$ over support $A$, then the second best standard is characterized by 
$$\bar{y} = \underset{z \in [0,Y]}{\arg\max} \displaystyle\int_{A} W(z,\alpha) dF(\alpha).$$
If each $W(z,\alpha)$ is concave over the relevant range, then the maximand above is also concave. Moreover, $$\bar{y} \in ( \underset{\alpha \in A}{\min}  \hspace{0.25cm} \bar{y}(\alpha), \underset{\alpha \in A}{\max} \hspace{0.25cm} \bar{y}(\alpha)).$$
Thus, if the smallest second best standard over the set $A$ exhibits a hawkish drift, then the standard associated with the case when $\alpha \in A$ is unknown also exhibits a hawkish drift.

We will now examine, given any performance standard, whether incumbents of a given ideology find it easier to meet the standard. In other words, we would like to know the relationship between $\theta_1$ and $\alpha$. Equation (\ref{thetaone}) gives us the relationship between any re-election standard $z$ and $\theta_1$. Differentiating equation (\ref{thetaone}) we get 
$$\frac{\partial \theta_1}{\partial \alpha} = \dfrac{-{z}^{2} \cdot \ln 2} {[(\alpha z
	+r) \cdot 2 \ln 2 -\alpha ^{2}]^2} 2 (\ln2 z - \alpha).$$
Given that the denominator is positive, 
\begin{equation}
	\frac{\partial \theta_1}{\partial \alpha} >(<) 0 \text{\hspace{0.5cm}if \hspace{0.5cm}} (\ln2 z - \alpha) <(>)0. \label{unknown hawk}
\end{equation} 

Thus we get,
\begin{proposition}
	Given any re-election standard $z$ such that $(\ln2 z - \alpha)>0$, the marginal (as well as average) quality of incumbent politicians who get re-elected is lower the more hawkish their ideology ( i.e., the greater the value of $\alpha)$.  
\end{proposition}
It is easy to see that the condition described in the Proposition above will always hold for $\alpha < 1$ for any $z \geq \frac{1}{\ln 2}$. For $\alpha >1$, this condition will be satisfied for a sufficiently high $z$. In all our simulations that we based our examples on, the optimal re-election standards under the first and the second best as well as the PBE satisfy this condition. Thus, we see that asymmetric information about ideology as well as ability can lead to
an endogenous bias in the electoral process. It essentially arises because
both ability and hawkishness leads to a higher $y,$ voters are unable to
distinguish the two and hence the optimal standard favors ideologically
hawkish politicians.}}

\section*{Discussion}\label{discussion section}
We have analyzed how the political process leads to an escalation of
conflict. The need for voters to screen high ability politicians leads them to set
standards which are more extreme than they would ideally like. We examine how the presence of rents from office as well as ideological divergence between politicians and voters affects policy choice. We show that both the second best as well as the incentive-compatible (i.e., PBE) standard vary non-monotonically with the incumbent ideology. Moreover, the PBE standard cannot be lower than the second best standard.

We also examine the effect of rents from office on the re-election standard. To satisfy incentive compatibility (i.e., under the PBE), politicians would be subject to a higher re-election standard as rents from office increase. This is not necessarily true for the second best. If the rents from office are high, it may be too costly to separate high ability politicians from low, and therefore, under the second best, the voters may choose to commit to a low re-election standard, thereby sacrificing on selection. However, such a strategy is not incentive compatible, and hence not a PBE. Further, when neither ideology nor ability is known, we find conditions under which there is an endogenous bias in the electoral process which favors the
re-election of hawks. Finally, given the divergence between the second best and the equilibrium of the political process (i.e., the PBE), we provide a rationale for how an international body like the UN can play a fruitful role in mitigating conflict by making people commit to treaties.

{\color{black}{It is worth noting that while conflict negotiation is a natural application of the environment we have modeled, any policy whose impact (and hence cost) will have to be managed the next period can give rise to the strategic interaction seen in our model. Leaders in office today will strategically commit to policies that are at least partially irreversible to tie the hands of their successor. Looking at what sort of distortions occur in this general environment is a promising area of future research.}}

In writing the model we made some assumptions which can be relaxed without qualitatively affecting the message of the paper. For instance, the probability and cost of conflict may depend on factors other than policy choice and incumbent ability, such as external shocks to the economy. Such a shock can be good (bad) making it less (more) costly to manage conflict. In Appendix \textbf{I.B} we discuss how such shocks can be modeled and their implications for our results. We consider different scenarios regarding whether the incumbent and/or the voter know the realized value of the shock before the policy is chosen. Under each scenario, we show that qualitative features of our results are unchanged. This gives us a general intuition: as long as a leader's ideal policy ($y$) is non-decreasing in ability, voters will associate higher ability with a higher observed $y$. Therefore, when leader's ability is not observable and there are rents from office, an incumbent will choose a hawkish policy to signal ability. 

In our model, we also assumed that a leader's ability matters only in managing the cost of the conflict, but not in managing other aspects of the economy. In Appendix \textbf{I.C} we modify our model to allow the leader's ability to determine both conflict and economic management and show that our results continue to hold as long as the conflict is over an economically valuable resource. This is because, due to his ability to manage the conflict better, a higher ability leader has a greater net resource at his disposal \textit{and} is able to generate a greater utility from better management of that resource.

There are several alternate reasons why conflict could give rise to
extremism. For instance, Glazer~\cite{glazer1998} shows that if policy reversal
is costly politicians may choose policies more extreme than the median voter
in order for rational voters to re-elect them as a costly policy reversal
will take place if the opposition comes to power. We would in that case
expect to see policy extremism in either direction i.e. a hawkish drift or a
dovish drift are both possible in such a scenario while in our model the
extremism is always in one direction. A related explanation which looks at
electoral concerns is by Hess and Orphanides~\cite{hess1995} where a politician of
unknown type goes to war when facing a sagging economy to see if he can
prove to be a better war leader. In that case if he tackles war well, that
becomes the salient issue and he is re-elected while if he is not, there is
nothing to lose as he would not have been re-elected anyway. One would
expect in that case that conflicts are endogenously created and vary with
the business cycle. In our model no such correlation would be seen which
gives rise to a testable hypothesis. Further, when leaders themselves do not
know their type as in Hess and Orphanides we would not see leaders who get
re-elected turn out to be hawks with a higher probability.

Another simple explanation of hawkishness is that conflict causes
a preference shift in the electorate so the median voter becomes more
hawkish leading to hawkish policies being implemented. However, assuming a
proportionate shift, all types become more hawkish than before; it will not
be the case that intrinsic hawks get re elected with a higher probability as
their distance from the median voter will not have changed. Similarly,
optimism bias would not imply that electoral concerns are what drives such
behavior and cannot explain the electoral advantage that intrinsic hawks
face.

We would want to extend this paper in two directions. The first is to
endogenize the candidate entry process using a citizen candidate model. We
want to see if the extremism in the political process can get mitigated by a
larger number of moderate people standing for election. We conjecture that
the answer is no. This is because more moderate people are likely to be
weeded out at a faster rate than more extreme people because of the bias in
the re-election process which reduces their incentive to stand as
candidates.\ Another line of work is to look at a potentially infinite
horizon model-as the low ability politicians get weeded out (for a given
ideology) at a higher rate than high ability ones, the distribution of
ability over time gets shifted towards a higher mean ability. However, this
will not lead to only the high ability people remaining in the long run. The
reason is that all types of politicians face a probability of death every
period, hence there will be a limiting distribution\cite{bandyopadhyay_2013}. We conjecture that the
policy will still be hawkish under this limiting distribution (and the
optimal standards set under this limiting distribution will still favor
hawks). These issues, along with a more rigorous examination of empirical
evidence of how conflict affects hawkishness are left for future
research.

\section*{Figures}
\paragraph*{Fig 1} Re-election Standards under Second Best and PBE, and FB with $\alpha =1$ 	
\paragraph*{Fig 2A} Re-election Standards under Second Best and PBE for $\alpha \geq 1$ 	
\paragraph*{Fig 2B} Re-election Standards under Second Best and PBE for $\alpha < 1$	

\section*{Supporting information}

% Include only the SI item label in the paragraph heading. Use the \nameref{label} command to cite SI items in the text.
{\color{black}{
\paragraph*{Appendix I}
\label{Appendix I} \textbf{Micro-Foundations and Theoretical Extensions}\\
\textbf{I.A} We construct a simple model to provide micro-foundations for the expected cost of insurgency as a function of $y$ and $\theta$. We assume that the members of $P$ have two choices - work, or join insurgency. The return to work are given by $\omega(y)$, which negatively depends on the land retained by $C$, i.e., $\omega (y) <0$. Each $P$ member also derives ideological benefits from joining the insurgency, which depends on his "type" $\delta$, denoted by $g(\delta)$ where $g(\delta)>0$, where $\delta$ is distributed according the CDF $F( )$.

A person of type $\delta$ will join insurgency if $g(\delta) \geq \omega(y)$ i.e. if 
$$\delta \geq g^{-1}(\omega(y)).$$ Hence, the mass of $P$ members joining insurgency is $	1 - F(g^{-1}(\omega(y)) )$. We assume that the potential cost of terrorist activity on $C$ members is proportional to the size of the insurgent group.

We also assume that the politician in office is able to counter the insurgency and thereby defuse the cost (without any loss of generality, to 0) with probability $P(\theta)$ which is an increasing function of $\theta$. Hence, the expected cost of insurgency is 
$$[1 - P(\theta)]\cdot [1 - F(g^{-1}(\omega(y)) )].$$

The specific example we used in our paper is $\omega(y) = 1 - \dfrac{y^2}{2}$, $F(\delta) \sim \mathcal{U} [0,1]$ and $1 -P(\theta) = \dfrac{1}{1+\theta}$. Hence we have
$$C(y, \theta) = \dfrac{y^2}{2(1 + \theta)}. $$}}

\textbf{I.B} In this part we check the robustness of our model to allowing for exogenous shocks to the cost of conflict to the society. A good shock means the expected cost of conflict is lower (either due to a lower probability of a conflict starting or a lower resultant damage from  the conflict) while a bad shock means the expected cost is higher. Let $\lambda (\geq 0)$ denote the parameter which determines the level of the shock, whose realized value is drawn from some probability distribution. Specifically, modify the voter's expected utility function in our baseline model by
$U = u(y) - \frac{1}{\lambda}\cdot c(y,\theta))$ which, with our specific functional form used in the paper, is
$$ y - \frac{1}{\lambda} \cdot \frac{y^2}{2(1+\theta)}.$$
There are three possibilities to consider:
\begin{enumerate}
	\item The value of $\lambda$ is unknown to the voter and the incumbent when choosing $y$.\\
	In this case the incumbent's ideal policy is $\arg\max y - E(\frac{1}{\lambda}) \cdot \frac{y^2}{2(1+\theta)}$, i.e., $$y = \frac{1}{E(1/ \lambda)} (1 + \theta).$$ As we can see, $y$ is monotonically increasing in $\theta$ making it qualitatively similar to the baseline case, and therefore generates the same strategic considerations as studied paper since a higher $y$ is a signal of higher incumbent ability. 
	\item Voter and incumbent know the value of $\lambda$ before choosing $y$.\\
	In this case the incumbent's choice is contingent on the realized value of $\lambda$, and is $y = \lambda (1+\theta)$. Hence, a good (bad) shock makes the policies more hawkish (dovish). Also, knowing this, the voters will set a more hawkish re-election standard in good times and a lower one in bad times.
	\item Incumbent knows the realized value of $\lambda$ but the voter does not.\\ 
	In this case, as before, the incumbent's ideal policy is $y = \lambda (1+\theta)$. However, since voter can  observe neither $\theta$ nor $\lambda$, screening is less precise. This is similar to the case of unknown incumbent ideology ($\alpha$) that we examined in the paper. Here too voters will apply Bayes rule to infer $\theta$ from the observed $y$ and based their re-election decision on this inference.
\end{enumerate}

\textbf{I.C} In this part we check the robustness of our model to allowing the incumbent's ability to affect both the cost of conflict as well as management of the economy. To this end, we modify the voter's payoff function to be
$$ U = \mu(\theta) \cdot [u(y) - \frac{c(y)}{\beta(\theta)}] $$
where $\beta(\theta)$ is the effectiveness of incumbent type $\theta$ in managing the conflict and $\mu(\theta)$ is the effectiveness of the leader in managing the economy (net of resources lost in conflict). Both $\mu(\cdot)$ and $\beta(\cdot)$ are strictly positive and weakly increasing in $\theta$. Therefore, maximizing the $U$ above is equivalent to maximizing the expression in the parenthesis as $\mu(\theta)$ acts as a shift parameter. In this case the results developed in the main section go through without any modification. This is because a higher ability leader has a greater net resource at his disposal, $u(y) - \frac{c(y)}{\beta(\theta)}$, and is able to generate a greater utility from better management of that resource.

An alternative formulation would be 
$$U = \mu(\theta)\cdot y(\theta) - \frac{c(y)}{\beta(\theta)}. $$
In this case, the optimal policy choice is given by the condition
$$ \frac{c'(y)}{u'(y)} = \mu(\theta)\cdot \beta(\theta). $$
Given the concavity of the $u(\cdot)$ function and the convexity of $c(\cdot)$, the LHS of the above equation is increasing in $y$ while the RHS is increasing in $\theta$, implying that a more able incumbent chooses a higher $y$. This leads to the same strategic considerations as studied in the baseline model since a higher $y$ acts as a signal of higher incumbent ability. 
{\color{black}{
\paragraph*{Appendix II} \label{Appendix II}
We provide here the values for the second best as well as the PBE re-election standard for various values of $\alpha$ and $r$. We also provide the corresponding values of $\theta_1$ and $\theta_2$ for the second best. These computations were performed using Mathematica as well as R. Code is available upon request.}} 
\begin{table}[H]
	\centering
\resizebox{8.3cm}{9.6cm}
	{\color{black}{
\begin{tabular}{|c|c|c|c|c|c|c|c|}
	\hline
	$a$ & $r$ & $W$ & $\bar{y}$ & $\theta_1$ & $\theta_2$ & SBE=PBE? & $\hat{y}$ \\
	\hline \hline
	0.9 & 0 & 0.792 & 1.559 & 0.484 & 0.732 & Yes & 1.559  \\
	\hline
	0.95 & 0 & 0.797 & 1.507 & 0.455 & 0.586 & Yes & 1.507 \\ 
	\hline
	1 & 0 & 0.799 & 1.443 & 0.443 & 0.443 & Yes & 1.443 \\ 
	\hline
	1.05 & 0 & 0.797 & 1.515 & 0.443 & 0.443 & Yes & 1.515 \\ 
	\hline
	1.1 & 0 & 0.791 & 1.587 & 0.443 & 0.443 & Yes & 1.587 \\ 
	\hline
	1.15 & 0 & 0.781 & 1.659 & 0.443 & 0.443 & Yes & 1.659 \\ 
	\hline
	0.9 & 0.05 & 0.793 & 1.644 & 0.429 & 0.826 & Yes & 1.644 \\ 
	\hline
	0.95 & 0.05 & 0.796 & 1.589 & 0.390 & 0.673 & Yes & 1.589 \\ 
	\hline
	1 & 0.05 & 0.797 & 1.528 & 0.363 & 0.528 & Yes & 1.528 \\ \hline
	1.05 & 0.05 & 0.795 & 1.456 & 0.353 & 0.387 & No & 1.595 \\ \hline
	1.1 & 0.05 & 0.789 & 1.496 & 0.360 & 0.360 & No & 1.665 \\ \hline
	1.15 & 0.05 & 0.780 & 1.572 & 0.367 & 0.367 & No & 1.734 \\ \hline
	0.9 & 0.1 & 0.790 & 1.713 & 0.388 & 0.904 & Yes & 1.713 \\ \hline
	0.95 & 0.1 & 0.791 & 1.655 & 0.341 & 0.742 & Yes & 1.655 \\ \hline
	1 & 0.1 & 0.791 & 1.594 & 0.306 & 0.594 & Yes & 1.594 \\ \hline
	1.05 & 0.1 & 0.789 & 1.526 & 0.284 & 0.453 & No & 1.658 \\ \hline
	1.1 & 0.1 & 0.784 & 1.446 & 0.278 & 0.315 & No & 1.726 \\ \hline
	1.15 & 0.1 & 0.775 & 1.485 & 0.291 & 0.291 & No & 1.795 \\ \hline
	0.9 & 0.15 & 0.785 & 1.773 & 0.353 & 0.970 & Yes & 1.773 \\ \hline
	0.95 & 0.15 & 0.785 & 1.710 & 0.301 & 0.800 & Yes & 1.710 \\ \hline
	1 & 0.15 & 0.784 & 1.647 & 0.261 & 0.647 & Yes & 1.647 \\ \hline
	1.05 & 0.15 & 0.781 & 1.580 & 0.231 & 0.505 & No & 1.710 \\ \hline
	1.1 & 0.15 & 0.776 & 1.506 & 0.214 & 0.369 & No & 1.778 \\ \hline
	1.15 & 0.15 & 0.768 & 1.443 & 0.217 & 0.255 & No & 1.846 \\ \hline
	0.9 & 0.2 & 0.778 & 1.815 & 0.318 & 1.000 & Yes & 1.815 \\ \hline
	0.95 & 0.2 & 0.776 & 1.757 & 0.267 & 0.849 & Yes & 1.757 \\ \hline
	1 & 0.2 & 0.774 & 1.691 & 0.222 & 0.691 & Yes & 1.691 \\ \hline
	1.05 & 0.2 & 0.771 & 1.623 & 0.188 & 0.546 & No & 1.756 \\ \hline
	1.1 & 0.2 & 0.766 & 1.552 & 0.164 & 0.411 & No & 1.823 \\ \hline
	1.15 & 0.2 & 0.759 & 1.472 & 0.154 & 0.280 & No & 1.891 \\ \hline
	0.9 & 0.25 & 0.769 & 1.844 & 0.283 & 1.000 & Yes & 1.844 \\ \hline
	0.95 & 0.25 & 0.766 & 1.797 & 0.236 & 0.892 & Yes & 1.797 \\ \hline
	1 & 0.25 & 0.763 & 1.728 & 0.188 & 0.728 & No & 1.730 \\ \hline
	1.05 & 0.25 & 0.759 & 1.659 & 0.150 & 0.580 & No & 1.797 \\ \hline
	1.1 & 0.25 & 0.754 & 1.589 & 0.122 & 0.444 & No & 1.864 \\ \hline
	1.15 & 0.25 & 0.748 & 1.513 & 0.105 & 0.315 & No & 1.931 \\ \hline
	0.9 & 0.3 & 0.759 & 1.872 & 0.251 & 1.000 & Yes & 1.872 \\ \hline
	0.95 & 0.3 & 0.755 & 1.831 & 0.207 & 0.928 & Yes & 1.831 \\ \hline
	1 & 0.3 & 0.750 & 1.759 & 0.157 & 0.759 & No & 1.768 \\ \hline
	1.05 & 0.3 & 0.746 & 1.689 & 0.116 & 0.608 & No & 1.834 \\ \hline
	1.1 & 0.3 & 0.741 & 1.618 & 0.085 & 0.471 & No & 1.901 \\ \hline
	1.15 & 0.3 & 0.735 & 1.545 & 0.063 & 0.343 & No & 1.968 \\ \hline
	0.9 & 0.35 & 0.748 & 1.898 & 0.222 & 1.000 & Yes & 1.898 \\ \hline
	0.95 & 0.35 & 0.742 & 1.861 & 0.180 & 0.959 & Yes & 1.861 \\ \hline
	1 & 0.35 & 0.737 & 1.785 & 0.127 & 0.785 & No & 1.802 \\ \hline
	1.05 & 0.35 & 0.732 & 1.713 & 0.084 & 0.632 & No & 1.869 \\ \hline
	1.1 & 0.35 & 0.727 & 1.642 & 0.051 & 0.493 & No & 1.936 \\ \hline
	1.15 & 0.35 & 0.721 & 1.570 & 0.026 & 0.365 & No & 2.003 \\ \hline
	0.9 & 0.35 & 0.748 & 1.898 & 0.222 & 1.000 & Yes & 1.898 \\ \hline
	0.95 & 0.4 & 0.728 & 1.886 & 0.154 & 0.985 & Yes & 1.886 \\ \hline
	1 & 0.4 & 0.723 & 1.807 & 0.099 & 0.807 & No & 1.834 \\ \hline
	1.05 & 0.4 & 0.717 & 1.000 & 0.000 & 0.000 & No & 1.901 \\ \hline
	1.1 & 0.4 & 0.712 & 1.000 & 0.000 & 0.000 & No & 1.968 \\ \hline
	1.15 & 0.4 & 0.706 & 1.000 & 0.000 & 0.000 & No & 2.035 \\ 
	\hline
\end{tabular}
}}
\end{table}
	\vspace{-0.5cm}
\newpage

\section*{Acknowledgments}\label{acknowledgement section}
We would like to thank participants at the first Political Economy workshop
at the University of Birmingham, seminar participants at the University of
Edinburgh, Ralph Bailey, Kalyan Chatterjee, Kaustav Das, Amrita Dhillon,
Jayasri Dutta, Jaideep Roy and Tomas Sj\"{o}str\"{o}m for helpful comments
on a very early version of this paper. We also thank two anonymous referees and the editor for their valuable comments which have helped improve the paper.

\section*{Data Availability Statement}\label{data availability}
All relevant data for this article, including codes (in Matlab 2020a, Mathematica 12, fortran 90), have been made available through the Aston University repository: \url{https://doi.org/10.17036/researchdata.aston.ac.uk.00000533}.

\nolinenumbers

% Either type in your references using
% \begin{thebibliography}{}
% \bibitem{}
% Text
% \end{thebibliography}
%
% or
%
% Compile your BiBTeX database using our plos2015.bst
% style file and paste the contents of your .bbl file
% here. See http://journals.plos.org/plosone/s/latex for 
% step-by-step instructions.
% 

\end{document}